\documentclass[aps,prl,twocolumn,showpacs,titlepage,superscriptaddress,10pt]{revtex4-1}
\newcommand{\Heff}{\mathcal{H}_{\mathrm{eff}}}
\newcommand{\bra}[1]{ \langle #1 |}
\newcommand{\ket}[1]{| #1 \rangle}
\newcommand{\bk}[2]{\left \langle #1 | #2 \right \rangle}
\newcommand{\aver}[1]{\left \langle #1 \right \rangle}
\usepackage{amssymb}
\usepackage[pdftex]{graphicx}
\usepackage{hyperref}
\hypersetup{colorlinks, linkcolor={blue},citecolor={blue}, urlcolor={blue}}

\begin{document}
\title{Experimental Width Shift Distribution: A Test of Nonorthogonality  for\\Local and Global Perturbations}
\author{J.-B.~Gros}
\affiliation{Laboratoire de Physique de la Mati\`ere Condens\'ee, CNRS, Universit\'e Nice Sophia Antipolis, UMR 7336, Parc Valrose, 06100 Nice, France.}
\author{U.~Kuhl}
\thanks{Corresponding author, email: ulrich.kuhl@unice.fr}
\affiliation{Laboratoire de Physique de la Mati\`ere Condens\'ee, CNRS, Universit\'e Nice Sophia Antipolis, UMR 7336, Parc Valrose, 06100 Nice, France.}
\author{O.~Legrand}
\affiliation{Laboratoire de Physique de la Mati\`ere Condens\'ee, CNRS, Universit\'e Nice Sophia Antipolis, UMR 7336, Parc Valrose, 06100 Nice, France.}
\author{F.~Mortessagne}
\affiliation{Laboratoire de Physique de la Mati\`ere Condens\'ee, CNRS, Universit\'e Nice Sophia Antipolis, UMR 7336, Parc Valrose, 06100 Nice, France.}
\author{E.~Richalot}
\affiliation{Universit\'e  Paris-Est, ESYCOM (EA 2552), UPEMLV, ESIEE-Paris, CNAM, 77454 Marne-la-Vall\'ee, France}
\author{D.~V.~Savin}
\affiliation{Department of Mathematics, Brunel University London, Uxbridge UB8 3PH, United Kingdom}
\published{26 November 2014 in \underline{Phys.\,Rev.\,Lett.\,\textbf{\textbf{113}},\,224101\,(2014)}}

\begin{abstract}
The change of resonance widths in an open system under a perturbation of its interior has been recently introduced by Fyodorov and Savin [Phys.\ Rev.\ Lett.\ \textbf{108}, 184101 (2012)] as a sensitive indicator of the nonorthogonality of resonance states. We experimentally study universal statistics of this quantity in weakly open two-dimensional microwave cavities and reverberation chambers realizing scalar and electromagnetic vector fields, respectively. We consider global as well as local perturbations, and also extend the theory to treat the latter case. The influence of the perturbation type on the width shift distribution is more pronounced for many-channel systems. We compare the theory to experimental results for one and two attached antennas and to numerical simulations with higher channel numbers, observing a good agreement in all cases.
\end{abstract}

\maketitle

The most general feature of open quantum or wave systems is the set of complex resonances. They manifest themselves in scattering through sharp energy variations of the observables and correspond to the complex poles of the $S$ matrix. Theoretically, the latter are given by the eigenvalues $\mathcal{E}_n=E_n-\frac{i}{2}\Gamma_n$ of the effective non-Hermitian Hamiltonian $\Heff$ of the open system \cite{verb85,soko89,fyod97,rott09}. The anti-Hermitian part of $\Heff$ originates from coupling between the internal (bound) and continuum states, giving rise to finite resonance widths $\Gamma_n>0$. The other key feature is that the eigenfunctions of $\Heff$ are nonorthogonal \cite{soko89,rott09}. Their nonorthogonality is crucial in many applications; it influences nuclear cross sections \cite{soko97ii}, features in decay laws of quantum chaotic systems \cite{savi97}, and yields excess quantum noise in open laser resonators \cite{schom00a}. For systems invariant under time reversal, like open microwave cavities studied below, the nonorthogonality is due to the complex wave functions, yielding the so-called phase rigidity \cite{brou03,kim05,bulg06b} and mode complexness \cite{savi06b,poli09b}.  Nonorthogonal mode patterns also appear in reverberant dissipative bodies \cite{lobk00a}, elastic plates \cite{xeri09}, optical microstructures \cite{wier08-11} and lossy random media \cite{bach14}.

Recently, such nonorthogonality was identified as the root cause for enhanced sensitivity to perturbations in open systems \cite{fyo12}, see also Ref.~\cite{Trefethen2005}. Consider the parametric motion of complex resonances under internal perturbations. This can be modeled by a Hermitian term $V$ added to $\Heff$, so $\Heff'=\Heff+V$. The complex energy shift $\delta\mathcal{E}_n = \mathcal{E}_n'-\mathcal{E}_n$ of the $n$th resonance is then given by perturbation theory for non-Hermitian operators \cite{fyo12,Kato}, yielding in the leading order $\delta\mathcal{E}_n = \bra{L_n}V\ket{R_n}$, where $\bra{L_n}$ and $\ket{R_n}$ are the left and right eigenfunctions of $\Heff$ corresponding to $\mathcal{E}_n$. They form a biorthogonal system; in particular, $\bk{L_n}{R_m}=\delta_{nm}$ but $U_{nm}\equiv\bk{L_n}{L_m}\neq\delta_{nm}$ in general. $U$ is known in nuclear physics as the Bell-Steinberger nonorthogonality matrix \cite{soko89,soko97ii}, see also \cite{chal98}. Crucially, a nonzero width shift $\delta\Gamma_n=-2\mathrm{Im}\,\delta\mathcal{E}_n$ is induced solely by the \emph{off-diagonal} elements of $U$ \cite{fyo12}
\begin{equation}\label{dgam}
 \delta\Gamma_n = i\sum_{m}(U_{nm}V_{mn}-V_{nm}U_{mn})\,,
\end{equation}
where $V_{nm}=\bra{R_n}V\ket{R_m}=V_{mn}^{*}$. It vanishes only if the resonance states were orthogonal (all $U_{m{\neq}n}=0$).

Note that the nonorthogonality measures studied in Refs.~\cite{schom00a,brou03,kim05,bulg06b,savi06b,poli09b} are related to the \emph{diagonal} elements $U_{nn}$. Those and the width shift contain complementary information on nonorthogonality. In particular, the off-diagonal elements $U_{nm}$ are parametrically stronger for weakly open systems, when the widths are small compared to the level spacing $\Delta$ ($\Gamma\ll\Delta$): then, $U_{n\neq m}\sim\frac{\Gamma}{\Delta}$ \cite{fyo12}, whereas $U_{nn}-1$ is of the order of $\bigl(\frac{\Gamma}{\Delta}\bigr)^2$ \cite{poli09b}. This leads to a higher sensitivity of $\delta\Gamma_n$ to nonorthogonality effects.

In this Letter, we report the first experimental study of the width shift statistics for fully chaotic systems using microwave cavities of different kinds. We consider both local and global perturbations and also investigate whether a different behavior occurs in the cases of scalar and vectorial fields.

\emph{Global versus local perturbations.---}
We consider only weakly open systems with time-reversal symmetry and model them by random matrix theory \cite{guhr98,fyod11ox}. The energy levels are then induced by the eigenvalues of a random matrix drawn from the Gaussian orthogonal ensemble (GOE). Those $N$ levels are coupled through the anti-Hermitian part of $\Heff$ to $M$ equivalent open channels \cite{verb85,soko89}, characterized by the same coupling  $\kappa\ll 1$. In such a regime, the resonance positions $E_n$ are given by those eigenvalues and reveal universal fluctuations on the local scale of $\Delta\sim\frac{1}{N}$ in the limit $N\gg1$. The resulting Gaussian statistics of the GOE eigenvectors (corresponding to the wave functions of the closed system) yields the well-known $\chi^2_M$ distribution
\begin{equation}\label{PTdis}
 P_M(\gamma) = \frac{1}{ 2^{M/2}\Gamma(M/2)}\gamma^{M/2-1}e^{-\gamma/2}
\end{equation}
for the rescaled resonance widths $\gamma_n=\pi\Gamma_n/(2\kappa\Delta)$ \cite{soko89,fyod97}.

\begin{figure}[t]
  \includegraphics[height=0.8\columnwidth]{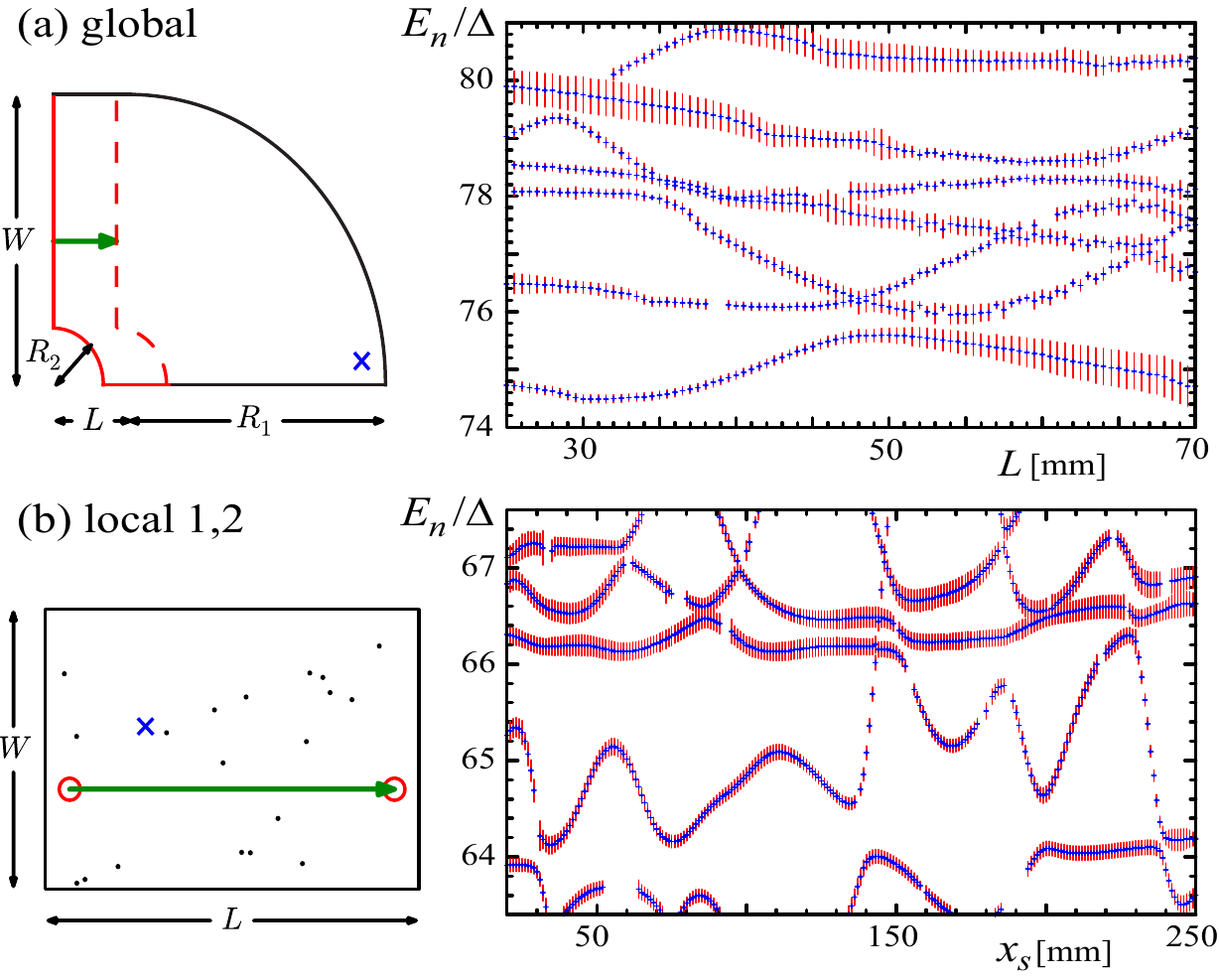}
  \caption{\label{fig:systems2D} (color online).
  The experimental setup (left) for two microwave cavities with one antenna (positioned at $\times$) together with the parametric dependence (right) of the Weyl normalized energies (+) and widths (red vertical lines): (a) Sinai stadium with a movable wall ($R_1{=}W{=}240$\,mm, $R_2{=}50$\,mm), $L$ ranging from 1.5 to 70.5\,mm in steps of 0.5\,mm. (b) Rectangular billiard ($L{=}340$\,mm, $W{=}240$\,mm) with 19 randomly placed scatterers (black dots) of radius $r_c{=}2.3$\,mm. One additional scatterer (red open circle) with radius $r_{p1}{=}2.3$\,mm or $r_{p2}{=}9.75$\,mm was moved along the line $x_s$ (green arrow) in steps of $\delta{r}{=}1$\,mm or 0.5\,mm, respectively. }
\end{figure}

To describe local and global perturbations on an equal footing, we follow Refs.~\cite{alei98,marc03} and represent the perturbation term as $V=\sum_{q=1}^{r}\alpha_q\ket{q}\bra{q}$. Its rank $r$ governs the transition between the local ($r$ small) and global ($r\gg1$) case. One can interpret $V$ as $r$ point scatterers characterized by the strength coefficients $\alpha_q$, where $q$ corresponds to their positions. For example, a single scatterer added to the (closed) system induces an energy shift $\delta{E}_n=\bra{n}V\ket{n}=\alpha\psi^2_n(q)$ for the $n$th level, with $\psi_n(q)=\bk{q}{n}$ being the wave function component at point $q$. However, moving the scatterer from point $q$ to $q'$, which we did in our experiment (see Fig.~\ref{fig:systems2D}), results in the shift $\delta{E}_n=\alpha(\psi_n(q)^2-\psi_n(q')^2)$. The latter is equivalent to a rank-2 perturbation with $V=\alpha(\ket{q}\bra{q}-\ket{q'}\bra{q'})$ \cite{marc03}. Generally, the variance of the energy shifts is given by $\mathrm{var}(\delta{E}_n)=\frac{2}{N^2}\mathrm{tr}(V^2)$, which sets up a scale for the parametric level dynamics \cite{fyod95b}. Importantly, the rescaled energy shifts (``level velocities'') $\sim\delta{E}_n/\sqrt{\mathrm{var}(\delta{E}_n)}$ acquire universal fluctuations of a distinct type in the case of local and global perturbations, being given by a $K_0$ distribution (for $r=2$) \cite{bar99d} and a Gaussian distribution, respectively. A gradual transition between the two occurs quickly as the perturbation rank $r$ grows \cite{marc03}.

In the same limit $\kappa\ll1$, Gaussian distributed wave functions result in the following representation for the rescaled width shifts (``width velocities'') \cite{fyo12}:
\begin{equation}\label{y_n}
 y_n \equiv \frac{\delta{\Gamma}_n}{2\kappa\sqrt{2\mathrm{var}(\delta{E}_n)}}
     = \frac{\sqrt{\gamma_n}}{\pi} \sum_{m\neq n}\frac{z_m v_m \Delta}{E_n-E_m}\,.
\end{equation}
Here, real $z_m$ are normally distributed random variables (stemming from coupling to the channels) whereas real $v_m=N\bra{m}V\ket{n}/\sqrt{\mathrm{Tr}(V^2)}$ are the normalized matrix elements ($m{\neq}n$) of the perturbation. These quantities are statistically independent of $E_n$ and $\gamma_n$, which is a result of separating independent fluctuations in spectra and in wave functions of weakly open chaotic systems.

To characterize the universal statistics of the width velocities (\ref{y_n}), we compute their probability distribution (at the spectrum center) $\mathcal{P}_M(y)=\Delta\langle\sum_{n=1}^N\delta(E_n)\delta(y-y_n)\rangle$, where $\langle\cdots\rangle$ denotes the ensemble average. Making use of the convolution theorem, it can be cast as follows \cite{fyo12}
\begin{equation}\label{shiftdist}
 \mathcal{P}_M(y) = \int_0^{\infty} \frac{d\gamma}{\sqrt{\gamma}}\,
                    P_M(\gamma)\, \phi\left( \frac{y}{\sqrt{\gamma}}\right)\,,
\end{equation}
where the function $\phi(y)$ is defined by
\begin{equation}\label{phi}
 \phi(y) = \int_{-\infty}^{\infty}\frac{d\omega}{2\pi} e^{i\omega y}
 \left\langle\prod_{m\neq n}\exp\left\{-i
 \frac{\omega z_m v_m}{\pi E_m/\Delta}
 \right\}\right\rangle\,.
\end{equation}

For global perturbations, the quantities $v_m$ become normally distributed random variables \cite{fyo12}, making the integration over $\{z_m, v_m\}$ straightforward. It results in the GOE average of certain spectral determinants, which was also derived in Ref.~\cite{fyo12}, with the explicit form of $\phi$ being
\begin{equation}\label{phigoe}
  \phi^{\mathrm{(gl)}}(y) = \frac{4 + y^2}{6(1+y^2)^{5/2}}\,.
\end{equation}
When substituted into Eq.~(\ref{shiftdist}), it leads to the distribution of the width velocities in the global case, $\mathcal{P}^{\mathrm{(gl)}}_M(y)$.

For local perturbations, $v_m$ have more complicated statistics. However, an exact result can be found in the particular case of $r$ equivalent scatterers (all $|\alpha_q|=\alpha$), which is of interest here. To this end, we first treat $v_m=\frac{N}{\sqrt{r}}(\vec\psi_m\cdot\vec\psi_n)$ as a scalar product of two $r$-dimensional vectors of the corresponding wave function components. It has a natural parametrization $v_m=\frac{\sqrt{\eta_n}}{\sqrt{r}}\nu_m$ in terms of the vector length $|\vec\psi_n|=\sqrt{\eta_n}$ and the projection $\nu_m$. The advantage of such a parametrization is that $\nu$ and $\eta$ are statistically independent \cite{poli09b}, with a normal and $\chi^2_r$ distribution [cf. Eq.~(\ref{PTdis})], respectively. Then, a Gaussian integration over $\{z_m, \nu_m\}$ in Eq.~(\ref{phi}) yields
\begin{equation}\label{philoc}
  \phi_r^{\mathrm{(loc)}}(y) = \aver{\frac{\sqrt{r}}{\sqrt{\eta}}\phi^{\mathrm{(gl)}}
  \left(\frac{\sqrt{r} y}{\sqrt{\eta}}\right)}_{\eta}\,,
\end{equation}
where $\phi^{\mathrm{(gl)}}$ is given by Eq.~(\ref{phigoe}) and the remaining average over $\eta$ is left at the end \cite{note1}. Combination of Eqs.~(\ref{shiftdist}) and (\ref{philoc}) solves the problem exactly at arbitrary rank $r$.

\begin{figure}[t]
  \includegraphics[width=\columnwidth]{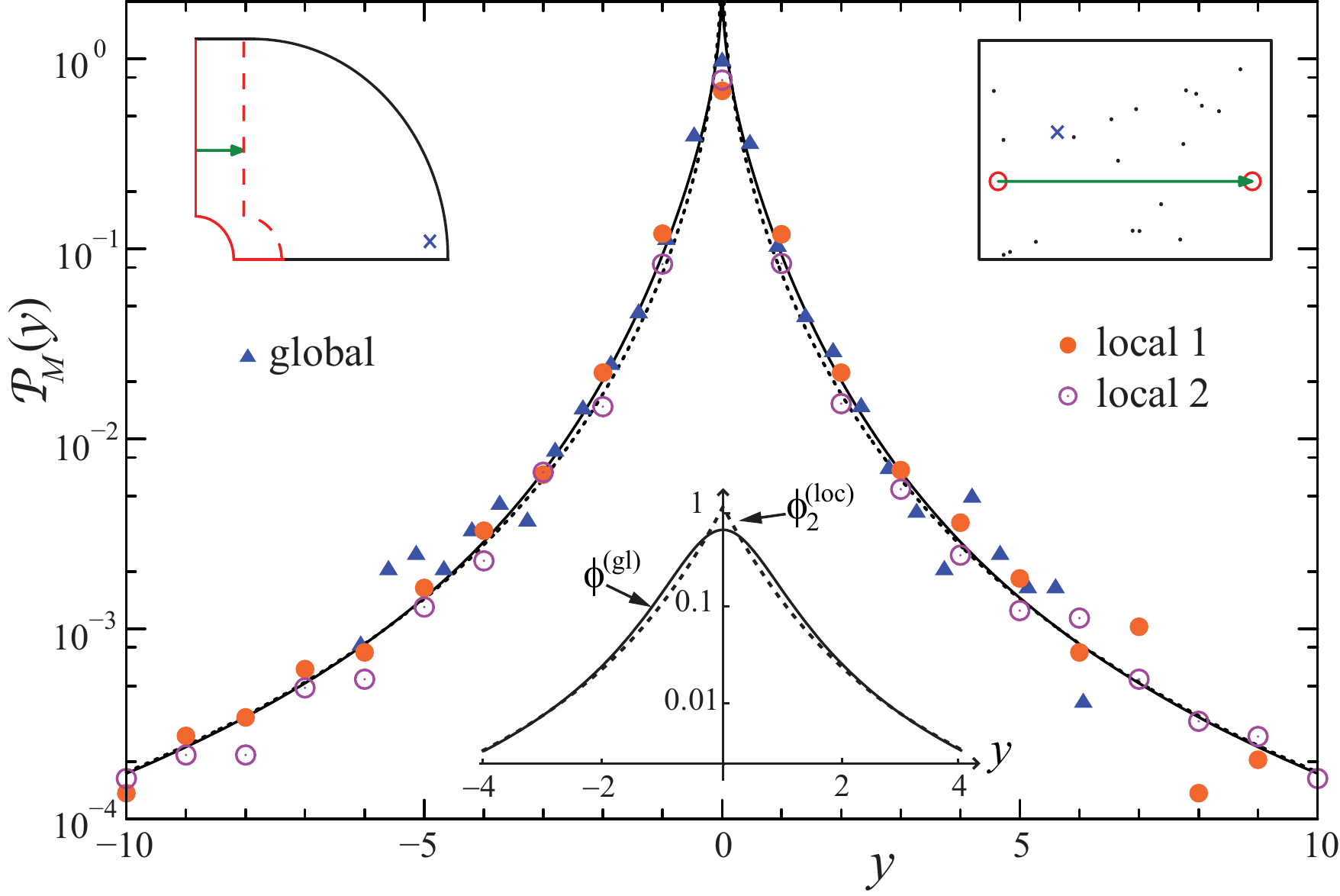}
  \caption{\label{fig:ExpShiftDist} (color online).
  Distribution of the experimental normalized width velocities $y$ for three systems corresponding to Fig.~\ref{fig:systems2D}: global ($\blacktriangle$), local 1, ($\bullet$) and local 2 ($\odot$). The solid (dashed) curve stands for the theoretical prediction for the global (local, $r{=}2$) perturbation with $M=1$. The lower inset shows the behavior of $\phi^{\mathrm{(gl)}}$ and $\phi_2^{\mathrm{(loc)}}$, see Eqs.~(\ref{phigoe}) and (\ref{philoc}).}
\end{figure}

Functional dependencies of $\phi_r^{\mathrm{(loc)}}(y)$ and $\phi^{\mathrm{(gl)}}(y)$ are the same in the tails and differ only in the bulk, but their difference diminishes quickly as $r$ grows. For small channel numbers, the difference becomes even less noticeable for the width velocity distribution $\mathcal{P}_{M}(y)$, e.g.~see Fig.~\ref{fig:ExpShiftDist}, due to the additional integration in Eq.~(\ref{shiftdist}) over the widths. Since the width distribution (\ref{PTdis}) tends to $\delta(\gamma-M)$ as $M\to\infty$, one has $\mathcal{P}_{M\gg1}(y)=\frac{1}{\sqrt{M}}\phi_r^{\mathrm{(loc)}}(\frac{y}{\sqrt{M}})$ as the limiting distribution of the width velocities in this case. Hence, many-channel systems turn out to be more sensitive to the impact of finite $r$ than their few channel analogues. We also mention the general  power-law decay $\mathcal{P}_M(y) \propto |y|^{-3}$ of the distribution at $|y|\gg1$, which can be linked to the linear level repulsion \cite{fyo12}. Such tails get exponentially suppressed in systems with rigid spectra without spectral fluctuations \cite{poli09b,savi13}.

\emph{Scalar experiments.---}
To investigate the statistics of the width velocity for scalar fields we use cylindrical (two-dimensional) microwave cavities, where the $z$ component of the electric field corresponds to the quantum wave function $\psi$ and the wave number $k^2$ to the energy $E$ \cite{stoe90-92}. Their heights are 8\,mm, leading to a cutoff frequency of $\nu_\mathrm{cut}=18.75$\,GHz and the frequency range used around 5\,GHz (wavelength 6\,cm). Figure~\ref{fig:systems2D} shows the three different systems. The first one is a chaotic Sinai-stadium billiard [see Fig.~\ref{fig:systems2D}(a)], which we will denote as the global perturbation. We used the range from the 50th to 100th resonance for the width velocity distribution. The second (third) system is a rectangular cavity with 19 scatterers, where an additional scatterer with the same (a larger) radius was moved [see Fig.~\ref{fig:systems2D}(b) and Ref.~\cite{bar99d} for further details], being denoted by local 1 (local 2). Again we took resonances from the 50th to 100th (85th) for the local 1 (2) case. All three systems are chaotic and in the ballistic regime, showing no level crossings experimentally.

The complex energies of the isolated resonances have been obtained by Lorentzian fitting. In all cases the energies and widths are normalized to the mean level spacing $\Delta$ by the Weyl formula $E_n/\Delta=\pi A (\nu_n/c)^2+P(\nu_n/c)$, where $\nu_n$, $A$, and $P$ are the eigenfrequency, area, and circumference, respectively. In case of the global perturbation, this unfolding also removes the global energy shift due to the area change.

The parametric dependence of the complex resonances for these systems is shown in Fig.~\ref{fig:systems2D}. Blue crosses indicate the resonance positions and the length of the red vertical lines corresponds to their widths. A distinct difference in the parametric level dynamics for the global and local perturbations is already visible here. This is further reflected in the level velocity distribution, which is a Gaussian distribution ($K_0$ distribution) in the global (local) case; both cases have been experimentally studied in Refs.~\cite{bar99c,bar99d}. Notably, such differences are much less pronounced for the width changes, as already discussed.

Obtaining the normalized width shifts (\ref{y_n}) from the measured data requires two parameters: the antenna coupling and the variance of $\delta{E}_n$. Both can be fixed in advance. The antenna coupling can be calculated by \cite{koeb10}
\begin{equation}\label{eq:defS11Kappa}
\kappa = \frac{\left|1-\langle S_{11} \rangle_{\nu,p}\right|^2}{1-\left|\langle S_{11} \rangle_{\nu,p}\right|^2},
\end{equation}
where $S_{11}$ is the complex reflection amplitude and special care has been taken to remove global phase shifts induced by the antennas. The average $\langle\cdots\rangle_{\nu,p}$ was performed over the whole investigated frequency range and for all parameters $p$, giving $\kappa=0.180$ (global), $\kappa=0.065$ (local 1), and $\kappa=0.098$ (local 2). We also took into account the absorption width $\Gamma_{n}^{(w)}$ due to the finite conductivity of the metallic walls, but neglected its variations, since $\Gamma_{n}^{(w)}$ as a function of the parameter induces much smaller changes than those due to the coupled antennas. Note that we do not assume that $\Gamma_{n}^{(w)}$ is the same for all resonances \cite{savi06b,pol10,kuh13}.

The experimental distributions of the width velocities are presented in Fig.~\ref{fig:ExpShiftDist}. After the normalization described above, there is no free parameter when comparing with the theoretical result (\ref{shiftdist}), as $M$ is fixed by the number of attached antennas, i.e.~$M=1$ here. In all cases, we find a good agreement with the corresponding theory. However, the amount of statistics is not sufficient to distinguish between the global and local perturbations in the width velocity distribution. At the only point where this would be possible statistically ($y$ close to 0), the experimental approximation of neglecting effects induced by absorption is no longer valid.

\begin{figure}[t]
  \includegraphics[width=\columnwidth]{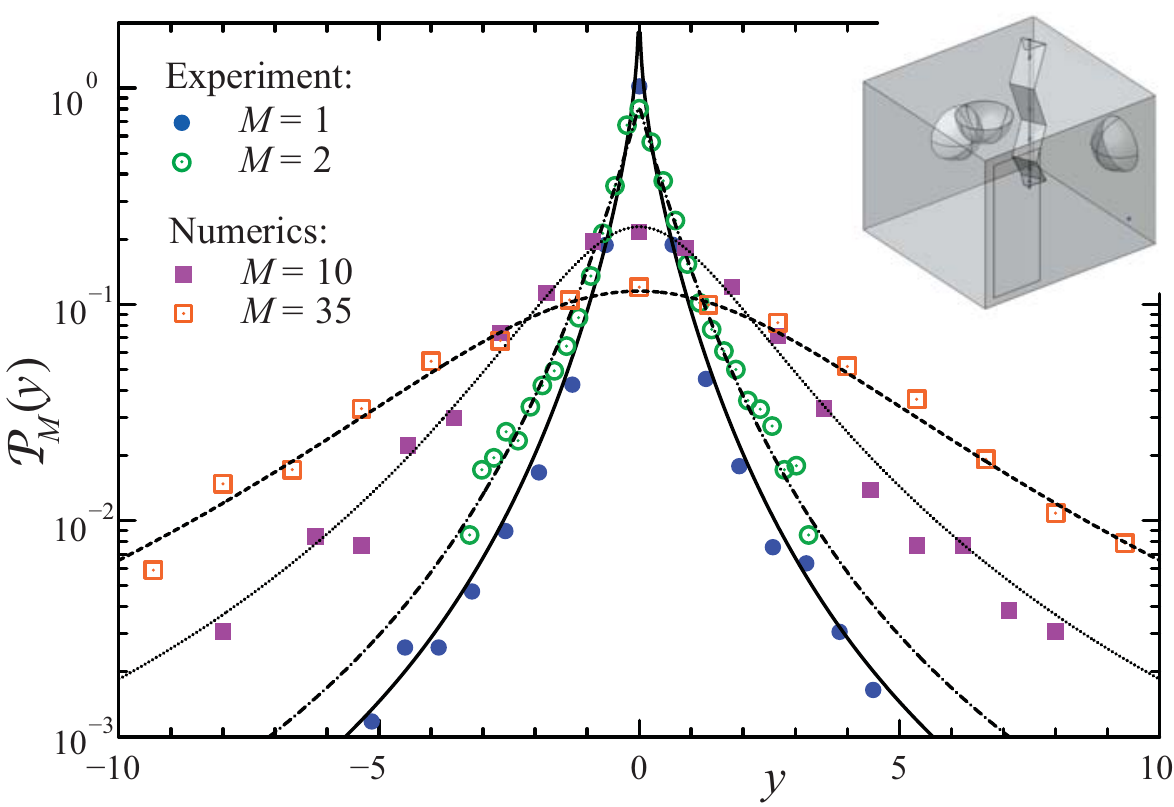}
  \caption{\label{fig:NumShiftDist} (color online).
  Distribution of the width velocities for several configurations of the chaotic reverberation chamber (shown in the inset) with rotating stirrer acting as a global perturbation. Experimental results correspond to $M$=1 and $M$=2 (circles), and numerical simulations to $M$=10 and $M$=35 (squares). The lines stand for the corresponding analytical results. For the normalization of the width shifts a fitted value of the coupling was used, yielding $\kappa_{M=1}=0.45$, $\kappa_{M=2}=0.16$, $\kappa_{M=10}=0.049$, and $\kappa_{M=35}=0.019$.}
\end{figure}

\emph{Vectorial electromagnetic cavities.---}
To support the universality of width shift fluctuations in the three-dimensional case of electromagnetic vector fields, we present experimental results as well as numerical simulations in a chaotic reverberation chamber (RC). We emphasize for this case the dependence on the channel number $M$ through various types of losses induced in the cavity either through antennas (experiments) or locally distributed Ohmic dissipation at walls (numerical simulations). The experiments were performed in a commercial RC of the approximate volume 19\,m$^3$  that was made chaotic by adding three metallic half spheres on the walls \cite{GrosWamot2014} (inset of Fig.~\ref{fig:NumShiftDist}). The parametric variation corresponds to the rotation of an asymmetric stirrer acting as a global perturbation. The measurements were performed via either one single dipole antenna connected in a wall ($M=1$) or between the latter antenna and a monopole antenna ($M=2$)  placed inside the cavity far from all walls. The mean quality factor was about 2500, corresponding to a moderate average modal overlap of $d=\langle\Gamma\rangle/\Delta \simeq 0.4{-}0.5$. By applying the harmonic inversion \cite{kuh08b}, we extracted around 70 resonance frequencies and their widths for each of 128 (90) positions of the rotating stirrer at $M=1$ ($M=2$). The resulting distributions of the width velocities are shown in Fig.~\ref{fig:NumShiftDist}, demonstrating a good agreement with the theoretical predictions (\ref{shiftdist}) and (\ref{phigoe}) in both cases \cite{note2}. Thus the width shift distribution, theoretically obtained for quantum chaotic systems, i.e.\ scalar fields, appears to be valid also for vectorial electromagnetic fields.

It is difficult to investigate the role of higher channel numbers experimentally, since the coupling of each antenna would have to be reduced, leading to too small signal-to-noise ratios for any practical
extraction of the complex resonances. Moreover, in such a case, all dissipative losses would become of the same order as those induced by antennas. Therefore, we performed numerical simulations using a finite-element method and calculated the resonances of two different configurations of the chaotic RC described in Refs.~\citep{GrosWamot2014,GrosCem2014}, where the coupling was mimicked by local absorption at the boundaries through Ohmic dissipative square patches scattered over the walls. By tuning the conductivity, size, and number of the patches, we can control the quality factor and hence the effective number of weak absorptive channels, which can be estimated as $M=2\langle \Gamma \rangle^2/\mathrm{var}(\Gamma)$ \citep{GrosPre2014}. With the coupling strength given by $\kappa \simeq\pi d/(2M)$, we obtained $M=10$, $d=0.34$, and $\kappa_{10}=0.05$ in one configuration investigated and $M=35$, $d=0.51$, and $\kappa_{35}=0.024$ in the other. For both configurations, the width velocity distributions are presented in Fig.~\ref{fig:NumShiftDist}, showing an excellent agreement with the theoretical predictions.

In conclusion, we experimentally verified the theoretical results for the width shift distribution \cite{fyo12} for global perturbations for scalar as well as for electromagnetic vector fields, supporting the universality of width shift statistics in weakly open chaotic systems. Additionally, we extended the theoretical approach to arbitrary rank perturbation, which was also found to be in good agreement with our experimental findings.

\begin{acknowledgments}
We acknowledge support by the Agence Nationale de la Recherche (ANR) via the project CAOREV (Reference: ANR-11-BS03-000).
\end{acknowledgments}


\begin{thebibliography}{99}
\bibitem[{not({\natexlab{b}})}]{note2}
\bibinfo{note}{Note that we fitted $\kappa$ using the distribution as we
  were not able to extract its value by Eq.~(\ref{eq:defS11Kappa}), due to a
  frequency dependent phase induced by the antennas.}
\end{thebibliography}
\end{document}